# Security in Wireless Sensor Networks: Issues and Challenges

Al-Sakib Khan Pathan
*Department of Computer Engg.*
*Kyung Hee University, Korea*
spathan@networking.khu.ac.kr

Hyung-Woo Lee
*Department of Software*
*Hanshin University, Korea*
hwlee@hs.ac.kr

Choong Seon Hong
*Department of Computer Engg.*
*Kyung Hee University, Korea*
cshong@khu.ac.kr

*Abstract* — Wireless Sensor Network (WSN) is an emerging technology that shows great promise for various futuristic applications both for mass public and military. The sensing technology combined with processing power and wireless communication makes it lucrative for being exploited in abundance in future. The inclusion of wireless communication technology also incurs various types of security threats. The intent of this paper is to investigate the security related issues and challenges in wireless sensor networks. We identify the security threats, review proposed security mechanisms for wireless sensor networks. We also discuss the holistic view of security for ensuring layered and robust security in wireless sensor networks.

## 1. Introduction

Wireless Sensor Networks (WSN) are emerging as both an important new tier in the IT ecosystem and a rich domain of active research involving hardware and system design, networking, distributed algorithms, programming models, data management, security and social factors [1], [2], [3]. The basic idea of sensor network is to disperse tiny sensing devices; which are capable of sensing some changes of incidents/parameters and communicating with other devices, over a specific geographic area for some specific purposes like target tracking, surveillance, environmental monitoring etc. Today's sensors can monitor temperature, pressure, humidity, soil makeup, vehicular movement, noise levels, lighting conditions, the presence or absence of certain kinds of objects or substances, mechanical stress levels on attached objects, and other properties [4]. In case of wireless sensor network, the communication among the sensors is done using wireless transceivers. The attractive features of the wireless sensor networks attracted many researchers to work on various issues related to these types of networks. However, while the routing strategies and wireless sensor network modeling are getting much preference, the security issues are yet to receive extensive focus. In this paper, we explore the security issues and challenges for next generation wireless sensor networks and discuss the crucial parameters that require extensive investigations.

Basically the major challenge for employing any efficient security scheme in wireless sensor networks is created by the size of sensors, consequently the processing power, memory and type of tasks expected from the sensors. We discuss these issues and challenges in this paper. To address the critical security issues in wireless sensor networks we talk about cryptography, steganography and other basics of network security and their applicability in Section 2. We explore various types of threats and attacks against wireless sensor network in Section 3. Section 4 reviews the related works and proposed schemes concerning security in WSN and also introduces the view of holistic security in WSN. Finally Section 5 concludes the paper delineating the research challenges and future trends toward the research in wireless sensor network security.

## 2. Feasibility of Basic Security Schemes in Wireless Sensor Networks

Security is a broadly used term encompassing the characteristics of authentication, integrity, privacy, nonrepudiation, and anti-playback [5]. The more the dependency on the information provided by the networks has been increased, the more the risk of secure transmission of information over the networks has increased. For the secure transmission of various types of information over networks, several cryptographic, steganographic and other techniques are used which are well known. In this section, we discuss the network security fundamentals and how the techniques are meant for wireless sensor networks.

**2.1 Cryptography**

The encryption-decryption techniques devised for the traditional wired networks are not feasible to be applied directly for the wireless networks and in particular for wireless sensor networks. WSNs consist of tiny sensors which really suffer from the lack of processing, memory and battery power [6], [7], [8], [9]. Applying any encryption scheme requires transmission of extra bits, hence extra processing, memory and battery power which are very important resources for the sensors' longevity. Applying the security mechanisms such as encryption could also increase delay, jitter and packet loss in wireless sensor networks [10]. Moreover, some critical questions arise when applying encryption schemes to WSNs like, how the keys are generated or disseminated. How the keys are managed, revoked, assigned to a new sensor added to the network or renewed for ensuring robust security for the

This work was supported by MIC and ITRC Project

network. As minimal (or no) human interaction for the sensors, is a fundamental feature of wireless sensor networks, it becomes an important issue how the keys could be modified time to time for encryption. Adoption of pre-loaded keys or embedded keys could not be an efficient solution.

## 2.2. Steganography

While cryptography aims at hiding the content of a message, steganography [11], [12] aims at hiding the existence of the message. Steganography is the art of covert communication by embedding a message into the multimedia data (image, sound, video, etc.) [13]. The main objective of steganography is to modify the carrier in a way that is not perceptible and hence, it looks just like ordinary. It hides the existence of the covert channel, and furthermore, in the case that we want to send a secret data without sender information or when we want to distribute secret data publicly, it is very useful. However, securing wireless sensor networks is not directly related to steganography and processing multimedia data (like audio, video) with the inadequate resources [14] of the sensors is difficult and an open research issue.

## 2.3 Physical Layer Secure Access

Physical layer secure access in wireless sensor networks could be provided by using frequency hopping. A dynamic combination of the parameters like hopping set (available frequencies for hopping), dwell time (time interval per hop) and hopping pattern (the sequence in which the frequencies from the available hopping set is used) could be used with a little expense of memory, processing and energy resources. Important points in physical layer secure access are the efficient design so that the hopping sequence is modified in less time than is required to discover it and for employing this both the sender and receiver should maintain a synchronized clock. A scheme as proposed in [15] could also be utilized which introduces secure physical layer access employing the singular vectors with the channel synthesized modulation.

## 3. Security Threats and Issues in Wireless Sensor Networks

Most of the threats and attacks against security in wireless networks are almost similar to their wired counterparts while some are exacerbated with the inclusion of wireless connectivity. In fact, wireless networks are usually more vulnerable to various security threats as the unguided transmission medium is more susceptible to security attacks than those of the guided transmission medium. The broadcast nature of the wireless communication is a simple candidate for eavesdropping. In most of the cases various security issues and threats related to those we consider for wireless ad hoc networks are also applicable for wireless sensor networks. These issues are well-enumerated in some past researches [16], [17], [18] and also a number of security schemes are already been proposed to fight against them. However, the security mechanisms devised for wireless ad hoc networks could not be applied directly for wireless sensor networks because of the architectural disparity of the two networks. While ad hoc networks are self-organizing, dynamic topology, peer to peer networks formed by a collection of mobile nodes and the centralized entity is absent [19]; the wireless sensor networks could have a command node or a base station (centralized entity, sometimes termed as sink).

The architectural aspect of wireless sensor network could make the employment of a security schemes little bit easier as the base stations or the centralized entities could be used extensively in this case. Nevertheless, the major challenge is induced by the constraint of resources of the tiny sensors. In many cases, sensors are expected to be deployed arbitrarily in the enemy territory (especially in military reconnaissance scenario) or over dangerous or hazardous areas. Therefore, even if the base station (sink) resides in the friendly or safe area, the sensor nodes need to be protected from being compromised.

### 3.1. Attacks in Wireless Sensor Networks

Attacks against wireless sensor networks could be broadly considered from two different levels of views. One is the attack against the security mechanisms and another is against the basic mechanisms (like routing mechanisms). Here we point out the major attacks in wireless sensor networks.

*3.1.1 Denial of Service*
Denial of Service (DoS) [20], [21] is produced by the unintentional failure of nodes or malicious action. The simplest DoS attack tries to exhaust the resources available to the victim node, by sending extra unnecessary packets and thus prevents legitimate network users from accessing services or resources to which they are entitled. DoS attack is meant not only for the adversary's attempt to subvert, disrupt, or destroy a network, but also for any event that diminishes a network's capability to provide a service. In wireless sensor networks, several types of DoS attacks in different layers might be performed. At physical layer the DoS attacks could be jamming and tampering, at link layer, collision, exhaustion, unfairness, at network layer, neglect and greed, homing, misdirection, black holes and at transport layer this attack could be performed by malicious flooding and desynchronization. The mechanisms to prevent DoS attacks include payment for network resources, pushback, strong authentication and identification of traffic.

*3.1.2 Attacks on Information in transit*
In a sensor network, sensors monitor the changes of specific parameters or values and report to the sink according to the requirement. While sending the report, the information in transit may be altered, spoofed, replayed again or vanished. As wireless communication is vulnerable to eavesdropping, any attacker can monitor the traffic flow and get into action to interrupt, intercept, modify or fabricate [22] packets thus, provide wrong information to the base stations or sinks. As sensor nodes typically have short range of transmission and scarce resource, an attacker with high processing power and larger communication range could attack several sensors at the same time to modify the actual information during transmission.

### 3.1.3 Sybil Attack

In many cases, the sensors in a wireless sensor network might need to work together to accomplish a task, hence they can use distribution of subtasks and redundancy of information. In such a situation, a node can pretend to be more than one node using the identities of other legitimate nodes (Figure 1). This type of attack where a node forges the identities of more than one node is the Sybil attack [23], [24]. Sybil attack tries to degrade the integrity of data, security and resource utilization that the distributed algorithm attempts to achieve. Sybil attack can be performed for attacking the distributed storage, routing mechanism, data aggregation, voting, fair resource allocation and misbehavior detection [24]. Basically, any peer-to-peer network (especially wireless ad hoc networks) is vulnerable to sybil attack. However, as WSNs can have some sort of base stations or gateways, this attack could be prevented using efficient protocols. Douceur [23] showed that, without a logically centralized authority, sybil attacks are always possible except under extreme and unrealistic assumptions of resource parity and coordination among entities. However, detection of sybil nodes in a network is not so easy. Newsome et. al. [24] used radio resource testing to detect the presence of sybil node(s) in sensor network and showed that the probability to detect the existence of a sybil node is:

$$Pr(detection) = 1 - \left(1 - \sum_{allS,M,G} \frac{\binom{s}{S}\binom{m}{M}\binom{g}{G}}{\binom{n}{c}} \frac{S-(m-M)}{c}\right)^r$$

Where, $n$ is the number of nodes in a neighbor set, $s$ is the number of sybil nodes, $m$ malicious nodes, $g$ number of good nodes, $c$ is the number of nodes that can be tested at a time by a node, of which $S$ are sybil nodes, $M$ are malicious (faulty) nodes, $G$ are good (correct) nodes and $r$ is the number of rounds to iterate the test.

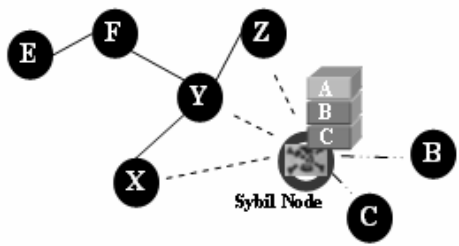

**Figure 1: Sybil Attack**

### 3.1.4 Blackhole/Sinkhole Attack

In this attack, a malicious node acts as a blackhole [25] to attract all the traffic in the sensor network. Especially in a flooding based protocol, the attacker listens to requests for routes then replies to the target nodes that it contains the high quality or shortest path to the base station. Once the malicious device has been able to insert itself between the communicating nodes (for example, sink and sensor node), it is able to do anything with the packets passing between them. In fact, this attack can affect even the nodes those are considerably far from the base stations. Figure 2 shows the conceptual view of a blackhole/sinkhole attack.

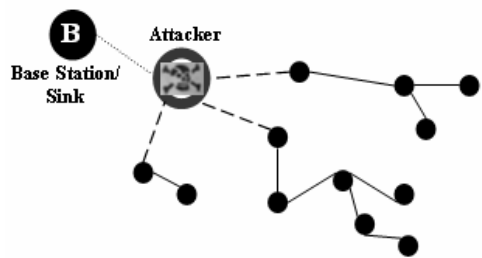

**Figure 2: Conceptual view of Blackhole Attack**

### 3.1.5 Hello Flood Attack

Hello Flood Attack is introduced in [26]. This attack uses HELLO packets as a weapon to convince the sensors in WSN. In this sort of attack an attacker with a high radio transmission (termed as a laptop-class attacker in [26]) range and processing power sends HELLO packets to a number of sensor nodes which are dispersed in a large area within a WSN. The sensors are thus persuaded that the adversary is their neighbor. As a consequence, while sending the information to the base station, the victim nodes try to go through the attacker as they know that it is their neighbor and are ultimately spoofed by the attacker.

### 3.1.7 Wormhole Attack

Wormhole attack [27] is a critical attack in which the attacker records the packets (or bits) at one location in the network and tunnels those to another location. The tunneling or retransmitting of bits could be done selectively. Wormhole attack is a significant threat to wireless sensor networks, because; this sort of attack does not require compromising a sensor in the network rather, it could be performed even at the initial phase when the sensors start to discover the neighboring information.

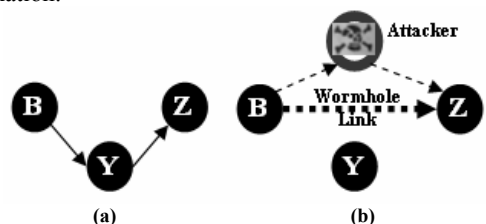

(a)          (b)
**Figure 3: Wormhole Attack**

Figure 3 (a and b) shows a situation where a wormhole attack takes place. When a node B (for example, the base station or any other sensor) broadcasts the routing request packet, the attacker receives this packet and replays it in its neighborhood. Each neighboring node receiving this replayed packet will consider itself to be in the range of Node B, and will mark this node as its parent. Hence, even if the victim nodes are multihop apart from B, attacker in this case convinces them that B is only a single hop away from them, thus creates a wormhole.

## 4. Proposed Security Schemes and Related Work

In the recent years, wireless sensor network security has been able to attract the attentions of a number of researchers around the world. In this section we review and map various security schemes proposed or implemented so far for wireless sensor networks.

### 4.1. Security Schemes for Wireless Sensor Networks

[26] gives an analysis of secure routing in wireless sensor networks. [34] studies how to design secure distributed sensor networks with multiple supply voltages to reduce the energy consumption on computation and therefore to extend the network's life time. [7] aims at increasing energy efficiency for key management in wireless sensor networks and uses Younis et. al. [36] network model for its application. Wood et al. [31] studies DoS attacks against different layers of sensor protocol stack. JAM [38] presents a mapping protocol which detects a jammed region in the sensor network and helps to avoid the faulty region to continue routing within the network, thus handles DoS attacks caused by jamming.

In [39] the authors show that wormholes those are so far considered harmful for WSN could effectively be used as a reactive defense mechanism for preventing jamming DoS attacks. Ye et. al. [33] presents a statistical en-route filtering (SEF) mechanism to detect injected false data in sensor network and focus mainly on how to filter false data using collective secret and thus preventing any single compromised node from breaking the entire system. SNEP & $\mu$TESLA [6] are two secure building blocks for providing data confidentiality, data freshness and broadcast authentication. TinySec [35] proposes a link layer security mechanism for sensor networks which uses an efficient symmetric key encryption protocol.

Newsome et. al. [24] proposes some defense mechanisms against sybil attack in sensor networks. Kulkarni et al. [28] analyzes the problem of assigning initial secrets to users in ad-hoc sensor networks to ensure authentication and privacy during their communication and points out possible ways of sharing the secrets. [40] presents a probabilistic secret sharing protocol to defend Hello flood attacks. The scheme uses a bidirectional verification technique and also introduces multi-path multi-base station routing if bidirectional verification is not sufficient to defend the attack.

Table 1: Summary of various security schemes for wireless sensor networks

| Security Schemes | Attacks Deterred | Network Architecture | Major Features |
|---|---|---|---|
| JAM [38] | DoS Attack (Jamming) | Traditional wireless sensor network | Avoidance of jammed region by using coalesced neighbor nodes |
| Wormhole based [39] | DoS Attack (Jamming) | Hybrid (mainly wireless partly wired) sensor network | Uses wormholes to avoid jamming |
| Statistical En-Route Filtering [33] | Information Spoofing | Large number of sensors, highly dense wireless sensor network | Detects and drops false reports during forwarding process |
| Radio Resource Testing, Random Key Pre-distribution etc. [24] | Sybil Attack | Traditional wireless sensor network | Uses radio resource, Random key pre-distribution, Registration procedure, Position verification and Code attestation for detecting sybil entity |
| Bidirectional Verification, Multi-path multi-base station routing [40] | Hello Flood Attack | Traditional wireless sensor network | Adopts probabilistic secret sharing, Uses bidirectional verification and multi-path multi-base station routing |
| On Communication Security [32] | Information or Data Spoofing | Traditional wireless sensor network | Efficient resource management, Protects the network even if part of the network is compromised |
| TIK [27] | Wormhole Attack, Information or Data Spoofing | Traditional wireless sensor network | Based on symmetric cryptography, Requires accurate time synchronization between all communicating parties, implements temporal leashes |
| Random Key Predistribution [29], [30], [41] | Data and information spoofing, Attacks in information in Transit | Traditional wireless sensor network | Provide resilience of the network, Protect the network even if part of the network is compromised, Provide authentication measures for sensor nodes |
| [42] | Data and Information Spoofing | Distributed Sensor Network, Large-scale wireless sensor network with dynamic nature | Suitable for large wireless sensor networks which allows addition and deletion of sensors, Resilient to sensor node capture |
| REWARD [43] | Blackhole attacks | Traditional wireless sensor network | Uses geographic routing, Takes advantage of the broadcast inter-radio behavior to watch neighbor transmissions and detect blackhole attacks |
| TinySec [35] | Data and Information spoofing, Message Replay Attack | Traditional wireless sensor network | Focuses on providing message authenticity, integrity and confidentiality, Works in the link layer |
| SNEP & $\mu$TESLA [6] | Data and Information Spoofing, Message Replay Attacks | Traditional wireless sensor network | Semantic security, Data authentication, Replay protection, Weak freshness, Low communication overhead |

REWARD [43] is a routing algorithm which fights against blackholes in the network. [32] proposes separate security schemes for data with various sensitivity levels and a location-based scheme for wireless sensor networks that protects the rest of the network, even when parts of the network are compromised. [27] implements symmetric key cryptographic algorithms with delayed key disclosure on motes to establish secure communication channels between a base station and sensors within its range. [41], [42], [29] and [30] propose key pre-distribution schemes, which target to improve the resilience of the network. In Table 1 we summarize various security schemes along with their main properties proposed so far for wireless sensor networks.

### 4.2. Holistic Security in Wireless Sensor Networks

A holistic approach [37] aims at improving the performance of wireless sensor networks with respect to security, longevity and connectivity under changing environmental conditions. The holistic approach of security concerns about involving all the layers for ensuring overall security in a network. For such a network, a single security solution for a single layer might not be an efficient solution rather employing a holistic approach could be the best option.

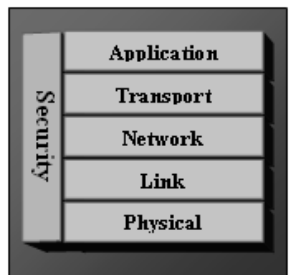

**Figure 4: Holistic view of Security in wireless sensor networks**

The holistic approach has some basic principles like, in a given network; security is to be ensured for all the layers of the protocol stack, the cost for ensuring security should not surpass the assessed security risk at a specific time, if there is no physical security ensured for the sensors, the security measures must be able to exhibit a graceful degradation if some of the sensors in the network are compromised, out of order or captured by the enemy and the security measures should be developed to work in a decentralized fashion. If security is not considered for all of the security layers, for example; if a sensor is somehow captured or jammed in the physical layer, the security for the overall network breaks despite the fact that, there are some efficient security mechanisms working in other layers. By building security layers as in the holistic approach, protection could be established for the overall network.

## 5. Conclusion

Most of the attacks against security in wireless sensor networks are caused by the insertion of false information by the compromised nodes within the network. For defending the inclusion of false reports by compromised nodes, a means is required for detecting false reports. However, developing such a detection mechanism and making it efficient represents a great research challenge. Again, ensuring holistic security in wireless sensor network is a major research issue. Many of today's proposed security schemes are based on specific network models. As there is a lack of combined effort to take a common model to ensure security for each layer, in future though the security mechanisms become well-established for each individual layer, combining all the mechanisms together for making them work in collaboration with each other will incur a hard research challenge. Even if holistic security could be ensured for wireless sensor networks, the cost-effectiveness and energy efficiency to employ such mechanisms could still pose great research challenge in the coming days.